\documentclass[runningheads]{llncs}
\usepackage[misc]{ifsym}
\usepackage[T1]{fontenc}
\usepackage{graphicx}
\usepackage{amsmath}
\begin{document}
\pagenumbering{gobble}
\title{A Measurement Study of LLM Inference Trade-offs Across Edge Continuum Hardware}
%
%
\author{
Maysam Khatib\inst{1} \and
Moysis Symeonides\inst{1}\textsuperscript{\Letter} \and
Demetris Trihinas\inst{2} \and
George Pallis\inst{1} \and
Marios D. Dikaiakos\inst{1}
}
\authorrunning{M. Khatib et al.}
\institute{
University of Cyprus, Nicosia, Cyprus\\
\email{\{mkhati01,msymeo03,pallis,mdd\}@ucy.ac.cy}
\and
University of Nicosia, Nicosia, Cyprus\\
\email{trihinas.d@unic.ac.cy}
}

\titlerunning{LLM Inference Trade-offs Across the Edge Continuum}

\maketitle              
\begin{abstract}

Large language models (LLMs) are increasingly used as backends for intelligent web services, but serving them across the edge continuum requires balancing quality, latency, model footprint, and energy. This paper presents a controlled measurement study of self-hosted LLM inference across edge and near-edge deployment nodes: an NVIDIA Jetson AGX Orin and a near-edge server with CPU-only and GPU-enabled inference modes. We evaluate multiple open-weight LLMs and quantization variants using a fixed question-answering workload, and compare them against GPT-4o as a cloud-hosted accuracy and latency reference. Our benchmarking pipeline reports accuracy, model footprint, per-token decoding latency, prefill latency, and overall execution energy. The results show that GPU-enabled server execution provides the lowest compute-side latency, while Jetson Orin shows lower measured energy, consistent with its lower platform power under our setup. CPU-only execution is consistently dominated in latency for our workload and shows higher measured energy. We also show that parameter count and downloaded weight-file size alone do not reliably predict observed accuracy or latency. Finally, using Pareto-frontier analysis, we study how deployment decisions may change under possible streamed-token delivery overheads, highlighting that compute-side inference metrics alone can lead to suboptimal placement for latency-sensitive interactive web services. 

\end{abstract}
\section{Introduction}
Large language models (LLMs) are rapidly moving from centralized cloud services to deployments that span the Edge continuum, from on-device execution to nearby edge servers and remote data centers~\cite{jiang2026edge}. This shift is driven by practical requirements such as tighter latency budgets, improved privacy, reduced bandwidth usage, and the need to operate under intermittent connectivity. At the same time, edge deployments introduce a new set of constraints. Hardware is heterogeneous, memory is limited, and power budgets are tight, especially on embedded accelerators and battery-powered nodes~\cite{Kasioulis2024}. As a result, deploying an LLM in real Edge-to-Cloud systems is no longer a single choice of model, but a sequence of decisions about which model to run, where to run it, and which trade-offs are acceptable for a given application and query~\cite{Morabito2025}.

In this setting, practitioners face competing objectives that rarely align. 
Higher accuracy is often associated with larger models, which in turn increase model footprint and can degrade responsiveness~\cite{Morabito2025}. 
Quantization can reduce memory needs and enable execution on edge devices, yet it can introduce overheads and quality shifts that depend on both the model and the serving stack~\cite{xiao2023smoothquant}. 
Moreover, the latency experienced by users is not only shaped by inference time. Even small network delays between the client and the server can offset compute-side latency advantages and change which model and deployment node are optimal for end-to-end performance~\cite{Morabito2025,Mohammed2020}. 
Lastly, energy usage further complicates these decisions, since higher-performance hardware may offer lower latency at higher power, while slower devices can be more energy-efficient~\cite{georgiou2025}.

These challenges reveal a gap not fully addressed by existing benchmarking, serving-system, quantization, and edge-LLM studies. Standardized benchmarks such as MLPerf Inference and MLPerf Power provide comparable performance and energy reporting across systems~\cite{reddi2020mlperf,mlcommons2025power}, while application-level benchmarks such as MMLU evaluate model reasoning across diverse subjects~\cite{hendrycks2021measuring}. However, these benchmarks are not designed to determine where an LLM-backed intelligent web service should execute across the edge continuum. Similarly, LLM serving systems such as Orca, vLLM, and Sarathi-Serve optimize batching, scheduling, memory management, and throughput-latency trade-offs within serving infrastructures~\cite{yu2022orca,kwon2023pagedattention,agrawal2024sarathi}, while quantization studies focus on compression methods for efficient inference~\cite{xiao2023smoothquant,lin2024awq}.  Lastly, Edge-oriented work such as CLONE further shows that LLM deployment feasibility depends on model design, system behavior, and hardware constraints~\cite{tian2025clone}, without examining the multi-objective trade-offs. 
In contrast, our work provides \textit{a controlled measurement-driven methodology for comparing model family, quantization, execution platform, latency, model footprint, and energy, together with a Pareto-based sensitivity analysis of streamed-token delivery overhead for intelligent web services.}
Specifically, this paper presents a controlled measurement study of LLM inference across selected edge and near-edge deployments. Rather than aiming to provide an exhaustive benchmark of all edge-LLM serving configurations, our goal is to quantify how model choice, quantization, execution platform, and streaming delivery overhead interact when selecting an LLM deployment under constraints on accuracy, responsiveness, model footprint, and energy. 

To this end, this paper makes four contributions: (i)~a containerized benchmarking pipeline for self-hosted LLM inference that records accuracy, model footprint, prefill and per-token decoding latency, and the overall execution energy; (ii)~we use this pipeline to compare edge and near-edge deployments, including on-device execution on Jetson Orin, near-edge GPU serving, and near-edge CPU serving, under a controlled interactive question-answering workload; (iii)~we quantify how model family, quantization level, and execution platform jointly affect the accuracy-latency-energy trade-off, showing that model size alone is not a reliable setup heuristic; and (iv)~we use Pareto-frontier analysis to identify dominated setups in the accuracy-latency space and to study how possible per-token streaming delivery overheads can affect deployment choices between local and server-side inference. 
The complete experimental configuration, including model identifiers and filenames, runtime/container versions, hardware settings, and prompt templates, is available in the versioned repository~\cite{khatib2026llmedgerepo}.

Our results highlight that no single off-the-shelf model is best across accuracy, latency, model footprint, and measured energy. GPU servers deliver the lowest per-token latency, while on-device execution on Jetson Orin shows lower measured energy, consistent with its lower platform power. Architectural differences and quantization overheads can outweigh parameter count when comparing practical latency and quality outcomes. 
Finally, we show that introducing possible per-token delivery overheads in a sensitivity analysis reshapes the accuracy-latency relation by reducing the latency advantage of server-side GPU inference and making on-device GPU inference more competitive for delay-sensitive settings, while server GPU remains preferable for higher-accuracy configurations. 
Together, these findings provide practical guidance for selecting LLMs and execution nodes under user and system constraints, and they can serve as a basis for preference-aware LLM routing in intelligent web services.

The rest of the paper is organized as follows. Sec.~\ref{sec:architecture} presents the benchmarking pipeline, setup, and methodology. Sec.~\ref{sec:observations} reports the results and Pareto-frontier analysis. Sec.~\ref{sec:rw} and Sec.~\ref{sec:conclusion} cover related work and the conclusion, respectively.

\section{Reference Benchmarking Architecture}
\label{sec:architecture}

\begin{figure}[t]
  \centering

    \centering
    \includegraphics[width=0.88\linewidth, trim={0cm 0.1cm 0.cm 0.05cm},
    clip]{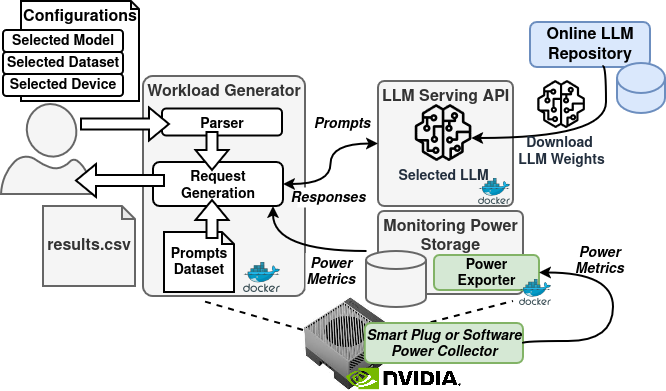}
    \vspace*{-0.5\baselineskip}
    \caption{Benchmarking Pipeline Overview}
        \label{fig:overview}
        \vspace*{-1.5\baselineskip}

\end{figure}
To enable efficient and fully automated benchmarking, we design a pipeline that orchestrates the experiment workflow from configuration to result export. Fig.~\ref{fig:overview} shows the overall architecture, which converts user-specified configurations into repeatable, measurable experiments that include accuracy, latency, and energy.

The workflow starts from a simple \texttt{configuration file} that defines the selected \textit{model}, \textit{dataset}, and \textit{target device}. A lightweight \textit{Parser} ingests these parameters and forwards them to the \textit{Request Generator}, which coordinates the execution. The Request Generator interacts with the \textit{LLM Serving API}, which is responsible for instantiating the chosen LLM. Upon initialization, the serving API downloads the required model weights from an \textit{Online LLM Repository} and loads them into the runtime before accepting inference requests. 
All components are deployed as containerized services to ensure portability and repeatability across heterogeneous edge platforms, with the serving module packaged as a multi-architecture image to support different CPU/GPU environments.

Once the model is ready, the Request Generator streams prompts from the selected dataset, normalizes them into a consistent request format, and submits them to the LLM Serving API. 
During execution, it records responses and timestamps to derive latency-related metrics and, for datasets that include expected outputs, computes task-level metrics such as accuracy.
In parallel, the monitoring subsystem collects power measurements through device-appropriate mechanisms, including external \textit{smart plugs} or \textit{software-exposed counters}. 
These metrics are exposed via a \textit{Power Exporter} and ingested by a \textit{Monitoring Power Storage} that stores time-aligned power metrics over the benchmark window.

After completion, the pipeline collects logs and monitoring traces, aggregates per-prompt and per-configuration summaries, and exports the results into a unified \texttt{results.csv} artifact. 
This output enables comparison across LLMs, datasets, and devices, allowing post-benchmark analysis. 

\subsection{Deployment and Implementation}

\noindent \textbf{Hardware.} In order to highlight how LLM design choices interact with hardware constraints, we evaluate two self-hosted deployment platforms and a cloud baseline. 
The first platform is an NVIDIA Jetson AGX Orin developer kit, an Arm-based edge system that integrates a 12-core Arm Cortex A78AE CPU with an NVIDIA Ampere GPU (2048 CUDA cores and 64 Tensor Cores) and 64 GB of 256-bit LPDDR5 memory, which makes it well suited for low-power on-device deployments while remaining sensitive to model size and runtime memory pressure.
The second platform is a near-edge server with an Intel Xeon Gold 6230 CPU, 96 GB RAM, and an NVIDIA T4 Tensor Core GPU with 2,560 CUDA cores, 320 Tensor Cores, and 16 GB of GDDR6 memory, providing higher throughput and lower inference latency than the Jetson-class device. 
Finally, we include GPT-4o as a cloud-hosted reference point to provide a reference accuracy baseline, noting that cloud executions do not expose device-level metrics comparable to our on-premise measurements. 

\noindent \textbf{Model Serving \& LLMs.}
For the LLM Serving API on edge nodes, we rely on the Ollama service~\cite{ollama} to host and query self-hosted models, while model weights are retrieved from the Hugging Face Hub~\cite{huggingface}, from which we select the evaluated set of open-weight LLMs.
Specifically, we focused on variations of \texttt{Llama 3.2} (Instruct model) with 1B and 3B parameters and F16, Q5\_K\_M, Q4\_K\_M quantization; \texttt{Mistral 7B} with Q5\_K\_M and Q4\_K\_M quantization; \texttt{TinyLlama 1.1B Chat v1.0} with Q5\_K\_M and Q4\_K\_M; and \texttt{Phi v2.0} with Q5\_K\_M and Q4\_K\_M quantized weights stored in GGUF format. We selected these quantization schemes because Q4\_K\_M and Q5\_K\_M represent a practical balance between resource usage and model quality. Q4\_K\_M provides strong compression (approximately 4.8 bits per weight~\cite{llamacpp_quantize_readme}), while Q5\_K\_M retains higher precision at increased memory cost. 
Both formats are widely adopted in practice. Specifically, Q5\_K\_M typically better preserves the model’s original performance and is considered preferable when hardware constraints permit, whereas Q4\_K\_M remains a reliable alternative for lower-memory environments. 
Compared to older legacy formats, the ``K'' variants reduce quantization error by using block-wise, importance-aware bit allocations, which can improve accuracy and, in some CPU-based settings, inference speed.


\noindent \textbf{Workload Generator.} 
We designed a workload generator as a Python script using modular abstractions that can operate across multiple datasets. 
For this study, we evaluate all models using the \textit{MMLU} (Massive Multitask Language Understanding) validation split, a widely used benchmark covering 57 academic subjects ranging from mathematics and computer science to the humanities~\cite{hendrycks2021measuring}.  
MMLU provides a controlled multiple-choice workload for comparing factual knowledge and reasoning accuracy across the evaluated configurations.  
We generate a total of 1531 prompts, issuing requests via the \texttt{Ollama} API for self-hosted models and the OpenAI API for the cloud baseline.
Each prompt is formatted into a strict multiple-choice question to ensure consistent evaluation across quantization levels, device types, and model families.
Since this workload asks each model to return a single multiple-choice option, it primarily evaluates short-output interactive inference; therefore, sustained long-form generation, long-context behavior, and stable inter-token streaming are outside the scope of our experiments.
All evaluations are initiated locally, with the workload generator deployed on the same node as the LLM on Orin or the GPU server.  
Thus, self-hosted models incur no WAN latency, as both request handling and inference are performed on the same device, ensuring that the measured latency captures only model computation and local serving overhead.
OpenAI API calls naturally include Internet latency, which is reported as part of the cloud baseline.
This setup allows us to isolate device-level performance and eliminates effects from network variability.
Moreover, we intentionally evaluate a sequential, single-request workload to model interactive LLM use cases in which users submit individual queries. This setting allows us to isolate the effect of model choice, quantization, execution device, and access latency without introducing batching and queueing effects. 
We note that we do not claim that the same ranking will hold under concurrent or multi-tenant serving. 
Parallel request streams, dynamic batching, and queue-aware scheduling can change both throughput and tail latency, and are treated as complementary dimensions for our future work.

\noindent \textbf{Reproducibility and Parametrization details.} All self-hosted experiments use the same serving stack, decoding parameters, and answer-extraction logic across devices and models. The answer generation is performed with the LLM temperature set to 0 and a fixed maximum output length sufficient to produce at least a single multiple-choice answer. 
Each prompt is formatted to request exactly one answer among the available options, and responses are parsed by extracting the first valid option label. 
Specifically, we use a zero-shot prompt, preserve the original option ordering, aggregate accuracy across all 1,531 questions, and count malformed responses as incorrect; the exact prompt and parser are available in~\cite{khatib2026llmedgerepo}. 
Before measurement, each model is loaded once and 10 warm-up prompts are executed to avoid including model download, initialization, and first-token warm-up effects in the reported latency. The measured interval begins immediately before request submission and ends when the complete model response is received by the workload generator.

\noindent \textbf{Recorded metrics.} For each trial, the workload generator records the LLM outputs and computes accuracy as the fraction of correctly predicted multiple-choice answers. 
It also records latency from request submission to response completion, together with prefill latency, decoding latency, and derived per-token latency metrics based on the data provided by the Ollama server and the number of input and generated tokens. 
Specifically, prefill latency per token is computed as prompt-evaluation duration divided by the number of input tokens, while decode latency per token is computed as generation duration divided by the number of generated tokens.
The resulting measurements are exported to a CSV file. 
Power metrics are collected separately through Prometheus~\cite{prometheus}, which acts as our Monitoring Power
Storage and periodically scrapes exporters running on the physical devices and stores the observed power values. Once the workload completes, the system queries monitoring storage to estimate the average power consumption \(\bar{P}\) during execution. It then computes energy per trial as \(E=\bar{P}\times T\), where \(T\) is the workload runtime, and stores it alongside the other captured metrics.

\noindent \textbf{Compute-side latency vs. client-observed streamed-token latency.}
We distinguish between compute-side latency and client-observed streamed-token latency. 
Compute-side latency per token is measured at the serving node and captures the time required by the model and serving stack to generate each output token. 
However, in interactive LLM applications, responses are often streamed token by token so that users can begin consuming the output before the full response is complete. 
In remote deployments, each generated token must also become available at the client, introducing additional transport, serialization, buffering, and client-side receipt costs.
For our sensitivity analysis (Sec.~\ref{sec:observations}), we approximate client-observed streamed-token latency as 
$L_{\text{stream}} = L_{\text{compute}} + d_{\text{delivery}}$, 
where $L_{\text{compute}}$ is the measured compute-side decoding latency per generated token, and $d_{\text{delivery}}$ is an effective per-token delivery overhead. 
The term $d_{\text{delivery}}$ abstracts the combined effects of pipelined delivery, buffering, persistent connections, and possible overlap between computation and communication into a single effective delivery penalty. 
We use this term to test how sensitive the accuracy-latency Pareto frontier is to remote-delivery effects.
The baseline self-hosted measurements report $L_{\mathrm{compute}}$, and, to study the sensitivity of deployment choices to delivery overhead, we add two fixed effective penalties, $d_{\mathrm{delivery}} = 30~\mathrm{ms}$ and $d_{\mathrm{delivery}} = 60~\mathrm{ms}$, to the server-side setups and recompute the accuracy-latency frontier. 
These values should be interpreted as controlled what-if parameters rather than measured network delays. 

\noindent \textbf{Power measurement.} In our evaluation, power measurements are interpreted as deployment-level indicators rather than as a calibrated cross-platform power audit. Specifically, on Jetson Orin, we use socket-level measurements~\cite{meross}, which capture total device draw during inference, including the board and surrounding platform components. On the server, we combine GPU power reported by \textit{nvidia-smi} with CPU package power reported by \textit{Intel RAPL}. This estimate excludes some platform-level components, including memory, storage, fans, motherboard losses, and power-supply losses. 
Hence, the server’s reported energy is a partial, lower‑bound estimate, while the Orin’s is a full‑system measurement. 
Moreover, we report average power over the execution interval and do not subtract idle power. Consequently, background platform consumption remains included. Because the Jetson measurement is socket-level whereas the server estimate covers CPU-package and GPU telemetry only, absolute cross-platform energy values should be interpreted as indicative rather than directly equivalent.
Because telemetry mechanisms differ across platforms, we focus primarily on robust qualitative trends rather than exact cross-platform energy equivalence. Moreover, because the OpenAI API does not expose hardware-level utilization or energy measurements, we exclude this model from the energy-related analysis.

\noindent \textbf{Variability.} Each trial is executed five times; we report mean values and use repeated runs to check whether the main ranking trends remain stable. Our goal is to compare central tendencies across deployment configurations; tail behavior such as p95/p99 latency, queueing delay, and request-level power transients are outside the scope of this sequential workload. Moreover, across repeated runs, the relative ranking of the main deployment configurations remained stable. 
The latter aspects are important for production serving systems, but they require separate experiments with parallel requests, batching, and tail-latency analysis.

\section{Observations}
\label{sec:observations}
Using our benchmarking pipeline, we evaluate multiple LLMs across five repeated runs and report average values. Next, we present the most notable results to examine key research questions on inference latency, model size, and energy usage for Edge-enabled LLM deployments.

\vspace*{0.5\baselineskip}
\noindent \textbf{RQ1: Which LLM setup provides the best trade-off between accuracy, latency, and model size for inference?}

\begin{figure*}[t]
  \centering

    \centering
    \includegraphics[width=0.95\linewidth, trim={0cm 2.cm 0.cm 1.6cm},
    clip]{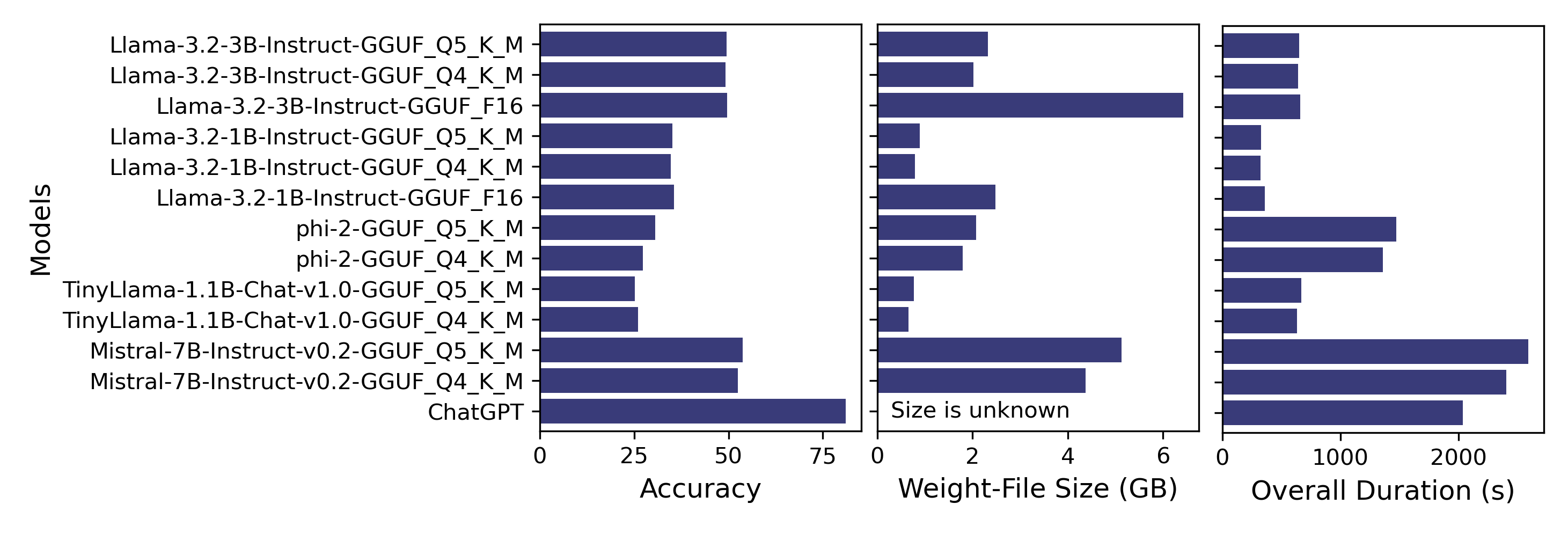}
    \vspace*{-.5\baselineskip}
    \caption{Performance of Different LLMs on Jetson Orin}
  \label{fig:llm_comparison}
      \vspace*{-2\baselineskip}

\end{figure*}
The results of our benchmarking are presented in Fig.~\ref{fig:llm_comparison}. Specifically, Fig.~\ref{fig:llm_comparison} (left) illustrates each model’s inference accuracy in relation to its size (middle) and observed latency (right) on the Jetson Orin platform. 
We also include \texttt{GPT-4o} as a cloud baseline.
As expected, \texttt{GPT-4o} achieves the highest accuracy, reaching approximately 80\% correct answers.
Among the self-hosted models, larger models such as \texttt{Mistral 7B} exhibit higher accuracy, albeit at the cost of significantly increased parameter counts and slower response times. Notably, the \texttt{Llama 3.2} F16 model, despite its larger weight-file size than the quantized variants, demonstrates slightly lower accuracy compared to smaller quantized models like \texttt{Mistral} Q5\_K\_M, and similar accuracy to its own quantized variants.
In terms of latency (Fig.~\ref{fig:llm_comparison}~(right)), however, the \texttt{Llama~3.2}~F16 model performs better than other 7B quantized models, suggesting that the observed latency is affected not only by model-file size, but also by model architecture, precision format, and serving-backend behavior.
These results show that parameter count and downloaded weight-file size alone do not reliably predict observed accuracy or latency. Model family, precision/quantization, serving stack, and execution platform jointly shape the measured outcome.
For instance, the quantized versions of the \texttt{Phi 2} are comparable in size to the quantized \texttt{Llama 3.2 3B} variants, yet the latter offer superior accuracy and latency performance.

\noindent \textit{\textbf{Observation:} Model size alone is an insufficient deployment heuristic. In our measurements, models with comparable footprints occupy different positions in the accuracy-latency space. This indicates that edge LLM selection should be based on measured configurations and workload-specific trade-offs rather than on model size alone. For intelligent services, this means that deployment decisions should consider model family, quantization, serving stack, and hardware together.}




\vspace*{0.5\baselineskip}
\noindent \textbf{RQ2: How does the choice of deployment platform influence LLM latency?}

\begin{figure}[t]

    \centering
    \includegraphics[width=1\linewidth, trim={0.45cm 0.95cm 0.3cm .4cm},
    clip]{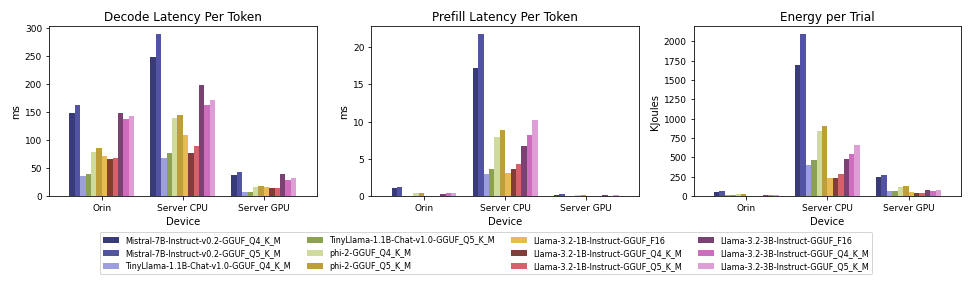}
    \vspace*{-2\baselineskip}
    \caption{Decode latency, prefill latency, and energy across deployment platforms}
  \label{fig:latencydevice}
\vspace*{-1.5\baselineskip}
\end{figure}

The left plot of Fig.~\ref{fig:latencydevice} reports average latency per generated token for a set of self-hosted LLMs, grouped by the device where inference runs. 
The x-axis shows three deployment options, \textit{Orin GPU}, \textit{Server CPU}, and \textit{Server GPU}. 
The y-axis shows milliseconds per token, so lower bars mean faster token generation and therefore a more responsive model during decoding. 
Each colored bar corresponds to a specific model and weight format, including multiple quantizations such as Q4 and Q5 and in some cases F16 precision.
A clear pattern is that the deployment platform dominates the latency. Across all models, Server GPU yields the lowest per-token latency, with bars clustered close to the bottom of the plot. 
Orin GPU is consistently slower than the server GPU but still remains lower than the server CPU deployment, indicating that GPU-accelerated inference on the Jetson provides reasonable token generation despite its more constrained compute and memory bandwidth. 
In contrast, Server CPU exhibits the highest per-token latency, with several models rising to the top of the chart. 
This means that even on a powerful server, moving inference from a GPU to CPU execution can increase token-level response time substantially.

Moreover, the middle plot of Fig.~\ref{fig:latencydevice} shows prefill latency, meaning how long the system needs to process the input prompt before it starts generating the output. 
The main observation is that the Server CPU has by far the highest prefill latency, especially for the larger models. For example, the 7B models and the larger quantized variants reach roughly 17–22 ms per input token, while smaller models remain lower but still clearly above the GPU results. This means that prompt processing on CPU becomes expensive as model size increases.
The Orin device has much lower prefill latency than the Server CPU, mostly around or below 1 ms per token. This suggests that Orin handles the prefill phase relatively efficiently for these models, despite being an edge device.
The Server GPU has the lowest prefill latency overall, with almost negligible values in the plot. This is expected because the prefill phase is highly parallelizable, and GPUs can process prompt tokens much more efficiently than CPUs.
Overall, it shows that prefill is strongly affected by the execution device and model size. GPUs are clearly the best option for fast prompt processing, CPUs suffer significantly for larger models, and Orin provides a strong edge alternative with much lower prefill latency than the Server CPU.
Within each device group, the multiple variants for the same base model show the effect of quantization. 
The Q4 and Q5 versions typically reduce memory traffic compared to F16, which usually results in faster token generation. 
On GPUs, Q4 and Q5 generally differ less, and all variants tend to converge to low values on the server GPU due to its higher throughput.

\noindent \textit{\textbf{Observation:} In this workload, execution platform is the largest observed factor for both decoding and prefill latency. Server GPU execution offers the lowest latency across the LLMs, while Orin GPU remains a competitive edge deployment option, offering substantially lower latency than CPU-only execution under tighter power and memory constraints. In contrast, Server CPU exhibits the highest latency in both phases, especially during prefill for larger models, making prompt processing and token generation slower without providing an accuracy benefit. Quantization affects performance within each device group, although its impact is less pronounced on the Server GPU where all variants converge to low latency. Thus, for the evaluated LLMs, server hardware, and our workload, CPU-only execution can be deprioritized when latency is the primary~objective.}

\noindent \textbf{RQ3: How do execution platform and model choice affect measured energy consumption?}

The right plot of Fig.~\ref{fig:latencydevice} shows the average energy consumed per trial for each evaluated model, grouped by the device where inference runs. The x-axis again shows the three deployment options \textit{Orin GPU}, \textit{Server CPU}, and \textit{Server GPU}. The y-axis shows energy per trial in kilojoules, so lower bars indicate that the system consumes less total energy over the benchmark trial.
First, we observe that the deployment platform strongly shapes energy needs. 
The CPU-only execution has the highest values by a large margin for most models, with several bars reaching multiple kilojoules per trial. 
This indicates that CPU-based inference is energy-expensive in our setup because longer execution times, especially during the prefill stage, increase the accumulated energy over the benchmark interval.
Furthermore, under our measurement setup, Orin shows the lowest measured energy per trial among the evaluated self-hosted setups, while the Server's consumption is also low, but it appears higher than Orin's. 
However, because the server measurement excludes memory, storage, fans, and PSU losses, the server’s true total energy would be higher than reported.
This suggests that the GPU-enabled server can provide high throughput and low latency, while an embedded edge device may still be attractive when lower platform power is prioritized. 
Additionally, within each device group, model choice and quantization still matter, but the device remains the dominant factor. Larger models and heavier variants tend to increase energy, especially on the CPU where higher compute costs directly extend generation time. Finally, lower-bit quantization variants, such as Q4 relative to Q5, reduce model size and memory traffic and may also reduce energy consumption, although the gains depend on the hardware and workload.


\noindent \textit{\textbf{Observation:} Energy consumption is governed by the interaction between generation time and platform power. 
Under our evaluated server configuration and specific workload, CPU-only inference shows higher measured energy because its substantially longer runtime outweighs its lower instantaneous device power.
Orin achieves lower measured energy values in this workload, plausibly because its lower platform power offsets its slower decoding speed relative to the GPU~server\footnote{We should note that, because Orin and server power are measured via different mechanisms, the absolute energy values should be interpreted cautiously, while the qualitative trend remains informative.}.}

\vspace*{0.5\baselineskip}
\noindent \textbf{RQ4: How can Pareto frontier analysis be used to identify and eliminate LLM-device configurations that are suboptimal in terms of accuracy and latency?} 

\begin{figure}[t]
    \centering
    \includegraphics[width=.99\linewidth, trim={0cm .5cm 0.cm 0.cm},
    clip]{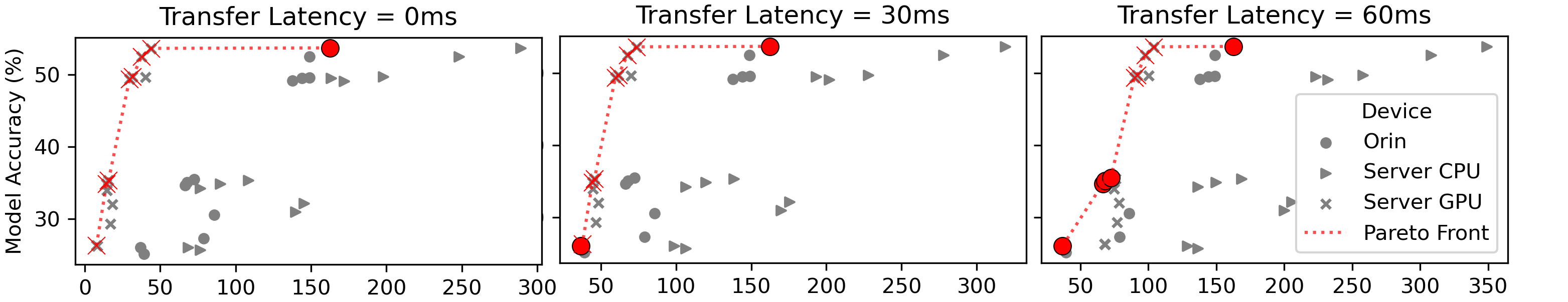}
        \vspace*{-1.\baselineskip}
    \caption{Accuracy-latency Pareto frontiers under different per-token delivery overheads.}
  \label{fig:paretodelay}
  \vspace*{-1.5\baselineskip}
\end{figure}

Our accuracy-latency trade-off analysis highlights that no single off-the-shelf LLM performs optimally across the evaluated configurations. Deploying a ``one-size-fits-all'' model can require practitioners to compromise between quality and responsiveness, with cost and energy requiring additional constraints or objectives.
Here, we examine how the Pareto front can be used as a tool for selecting the most efficient configuration within a deployment. Specifically, Fig.~\ref{fig:paretodelay} (left) plots each evaluated configuration as a point in the accuracy versus latency space, where the y-axis shows the percentage of correct responses and the x-axis shows average latency per generated token in milliseconds.
Marker shapes encode the deployment platform, with "diamond" points for Orin GPU, "x" for GPU-enabled deployment, and the "triangle" for CPU-enabled deployment. 
The dotted red curve highlights the Pareto front, that is, the set of configurations not dominated by any other option, meaning that no other point achieves both higher accuracy and lower latency simultaneously.



GPU-enabled Server deployments concentrate on the far left, indicating the lowest per-token latency, typically within a few milliseconds, while spanning a moderate range of accuracies. 
Additionally, Orin points are in the low latency region but with higher latency than the server GPU for many models, and they cover a wider spread in accuracy. By contrast, CPU-enabled deployment points appear further to the right, often at tens of milliseconds per token, showing that CPU execution incurs a large responsiveness penalty without a corresponding accuracy gain. 
Within our evaluated accuracy–latency configuration space, this makes all CPU-only trials dominated, since similar accuracies are achievable on GPU deployments at substantially lower latency. 
Moreover, the Pareto front starts from very low latency with lower accuracy, then climbs toward higher accuracy as latency increases, and ends at a high accuracy point that requires noticeably more latency than the fastest options. 
On the Pareto front itself, almost all points come from the GPU-enabled server, and only one Orin point provides high accuracy and relatively low latency.


\noindent \textit{\textbf{Observation:} There is no universally best LLM, so selection should be Pareto-driven and dominated options, such as TinyLlama, Phi, or CPU-enabled deployments, can be deprioritized for the evaluated workload when optimizing only accuracy and per-token latency.}

\vspace*{0.5\baselineskip}

\noindent \textbf{RQ5: How does streamed-token delivery delay affect the accuracy-latency Pareto frontier between on-device inference and edge-server LLM serving?} 

Continuing our analysis, we study how potential streaming delivery overheads can change the apparent performance of different deployment options. The previous measurements for the edge server reflect compute-side decoding latency and exclude the additional time required for generated tokens to become visible to a remote client. In many interactive LLM-enabled web services, responses are streamed token by token so that users can start reading before the full response is complete. Under this serving model, generated tokens may experience effective delivery overheads caused by transport, serialization, buffering, server-side flushing, and client-side receipt, although these costs can overlap with computation in real deployments.
To approximate this effect, we add two fixed effective per-token delivery overheads, namely 30 ms and 60 ms, to the server-side per-token latency before recomputing the Pareto frontier. These values are not intended to represent a fixed network round-trip delay for every token, nor measurements from a specific production edge network. Instead, they are used as controlled sensitivity parameters that represent moderate and high effective streaming overheads. This choice is consistent with prior cloud-edge latency measurements showing that latency differences between cloud and edge deployments often fall within the tens-of-milliseconds range~\cite{charyyev2020latency}. The sensitivity analysis is also motivated by recent cloud-edge LLM inference studies, which identify communication overhead as a key contributor to inference latency~\cite{Mingjin2025}.
We do not apply this adjustment to Orin configuration, since it represents on-device inference with no additional access-network hop.

Introducing effective per-token delivery overhead changes the shape of the Pareto frontier, but it does not remove the server GPU from the efficient set, as shown in Fig.~\ref{fig:paretodelay}. As the added overhead increases to 30 ms and 60 ms, the server GPU points move to the right, reducing their advantage in terms of client-observed token arrival delay. Under these overheads, local Orin execution becomes increasingly relevant in the low-delay region because it does not include the modeled remote delivery penalty.
However, server GPU setups remain on the Pareto frontier for higher-accuracy models, showing that the best deployment depends on the target accuracy-delay trade-off.
Therefore, this analysis should be viewed as a controlled sensitivity study of client-observed token arrival delay, rather than a measurement of a specific production network.

\noindent \textit{\textbf{Observation:} For streaming LLM services, deployment decisions should consider client-observed token arrival latency, not only compute-side inference time. As delivery overhead increases, server GPU setups lose part of their latency advantage, making local GPU-enabled devices more attractive for latency-sensitive workloads. However, server GPU remains Pareto-efficient for higher-accuracy configurations, so the best deployment depends on the accuracy-delay trade-off.
}

  \vspace*{-1.\baselineskip}
\section{Related Work}
\label{sec:rw}

Standardized benchmarking efforts have shaped how inference performance and efficiency are reported across platforms. MLPerf Inference defines representative tasks together with a common load generator, accuracy checking, and prescriptive run and submission rules that make results comparable across diverse hardware and software stacks~\cite{reddi2020mlperf}. MLCommons further complements this direction with the MLPerf Power methodology, which specifies measurement and reporting procedures to enable consistent power and energy comparisons across systems~\cite{mlcommons2025power}. In addition, MMLU is widely used as an application-level benchmark for evaluating the reasoning and knowledge capabilities of language models across diverse academic subjects~\cite{hendrycks2021measuring}.  Unlike such benchmarks, we do not introduce new tasks, but provide a methodology for edge LLM serving that reports accuracy, model footprint, compute-side latency, and measured energy, while also exploring possible streaming-delivery overheads through sensitivity analysis.

Focusing on making LLM inference feasible under tight memory limits, another large body of work targets LLM compression and quantization. 
Specifically, SmoothQuant proposes a training-free transformation that smooths the activation outliers by offline migrating the quantization burden from activations to weights through a mathematically equivalent transformation, while causing only minimal accuracy loss~\cite{xiao2023smoothquant}. 
In addition, AWQ~\cite{lin2024awq} targets low-bit weight-only quantization by identifying salient channels using activation statistics, protecting a fraction of weights, and pairing the method with a deployment-oriented runtime to accelerate on-device inference. 
Rather than proposing a new quantizer, our study treats quantization as a trial factor in an end-to-end pipeline and captures its effect on the underlying metrics.

LLM serving systems have improved throughput by rethinking scheduling around model compute and memory behavior and infrastructure capabilities. 
For instance, Orca~\cite{yu2022orca} introduces iteration-level scheduling and selective batching to balance latency and throughput for large transformers. vLLM~\cite{kwon2023pagedattention} further improves efficiency with PagedAttention, reducing KV-cache waste and increasing effective batch sizes, while Sarathi-Serve~\cite{agrawal2024sarathi} addresses the throughput-latency tradeoff through chunked prefills and stall-free scheduling, enabling larger batches with lower latency impact and fewer pipeline bubbles.
Recent edge-oriented work shows that deployment feasibility depends jointly on model design, system behavior, and hardware limits. For example, CLONE~\cite{tian2025clone} quantifies the challenges of running LLMs on edge devices and proposes an energy- and latency-aware algorithm-hardware co-design. 
Unlike prior work, we present a measurement-driven methodology for quantifying accuracy and system-level trade-offs across a self-hosted edge continuum, enabling principled local-versus-remote selection.

\vspace*{-.5\baselineskip}
\section{Conclusion, Limitations \& Future Work}
\label{sec:conclusion}
\vspace*{-.5\baselineskip}


This paper presented a controlled measurement study of LLM inference across edge and near-edge platforms, focusing on the trade-offs that arise when intelligent web services must choose between local, near-edge, and cloud-hosted execution. Using a containerized benchmarking pipeline, we evaluated multiple open-weight LLMs and quantization variants on Jetson Orin and on a near-edge server in GPU-enabled and CPU-only modes, while using GPT-4o as a cloud-hosted reference. The results show that LLM deployment is inherently multi-objective: accuracy, latency, model size, and measured energy do not always improve together. 
GPU-enabled server execution provides the lowest compute-side latency, Jetson Orin shows lower measured energy, and CPU-only execution is dominated in the accuracy-latency space for the studied workload. 
The results further show that parameter count alone does not reliably predict deployment performance, highlighting the need for empirical evaluation across model architectures and quantization settings. 
Our Pareto-frontier analysis shows that deployment choices depend on whether latency is measured only at the compute side or at the client-observed streamed-token level. When per-token streaming delivery overhead is included, some server-side configurations become less attractive despite faster decoding. This finding is particularly important for latency-sensitive interactive web services, where user-perceived responsiveness depends not only on token generation speed but also on how quickly generated tokens arrive at the client. Overall, our study provides evidence that preference-aware LLM routing across the edge continuum can benefit from jointly considering model quality, compute-side latency, model size, and energy.

\noindent \textbf{Limitations.} This study focuses on sequential interactive question answering and does not evaluate concurrent request streams, batching, queueing, or tail latency. Therefore, the reported rankings should not be generalized directly to high-throughput multi-tenant serving. Moreover, the workload is based on MMLU, a multiple-choice benchmark that captures factual and reasoning accuracy but does not cover long-context dialogue, retrieval-augmented generation, tool use, or multimodal tasks. In addition, because the MMLU workload produces short multiple-choice responses, the decoding and streaming results should not be generalized to long-form generation without additional experiments. Energy measurements rely on platform-specific telemetry with different measurement boundaries across devices; therefore, they should be interpreted as indicative deployment-level measurements rather than fully calibrated cross-platform comparisons.  Finally, the streamed-token delivery analysis uses controlled delay injection rather than measurements over production edge networks. These limitations motivate the broader evaluation planned in future work.

\noindent \textbf{Future work.} We plan to extend this controlled study along several directions. First, we aim to evaluate a broader set of edge-continuum configurations, including additional accelerators, near-edge servers, serving stacks, and memory systems, to assess how robust the observed deployment rankings are across different hardware and software environments. Second, we plan to expand the workload beyond sequential interactive question answering to include concurrent serving, long-context dialogue, retrieval-augmented generation, and agentic tool-use scenarios. 
Moreover, we intend to replace the controlled delay injection with measurements over real edge networks, capturing latency variability and its impact on model placement and routing decisions. 
Lastly, our analysis can serve as a foundation for prompt routing systems that combine user preferences with edge-to-cloud deployment performance to optimize end-to-end execution.

\vspace*{.5\baselineskip}
\scriptsize{\noindent\textbf{Acknowledgment.} 
This work is supported by the EU Commission through the AI-DAPT project (HORIZON-CL4-2023-HUMAN-01-01, GA: 101135826). Language refinements were performed at the sentence level using ChatGPT. All original content and ideas are solely those of the authors.}

%
%
%
\bibliographystyle{splncs04}
\bibliography{bibliography}

@INPROCEEDINGS{reddi2020mlperf,
  author={Vijay Janapa Reddi and others},
  booktitle={2020 ACM/IEEE 47th Annual International Symposium on Computer Architecture (ISCA)}, 
  title={MLPerf Inference Benchmark}, 
  year={2020},
  volume={},
  number={},
  pages={446-459},
  publisher = {IEEE},
  address = {Piscataway, NJ}}

@ARTICLE{Mingjin2025,
  author={Zhang, Mingjin and Shen, Xiaoming and Cao, Jiannong and Cui, Zeyang and Jiang, Shan},
  journal={IEEE Internet of Things Journal}, 
  title={EdgeShard: Efficient LLM Inference via Collaborative Edge Computing}, 
  year={2025},
  volume={12},
  number={10},
  pages={13119-13131}}

@INPROCEEDINGS{charyyev2020latency,
  author={Charyyev, Batyr and Arslan, Engin and Gunes, Mehmet Hadi},
  booktitle={GLOBECOM 2020 - 2020 IEEE Global Communications Conference}, 
  title={Latency Comparison of Cloud Datacenters and Edge Servers}, 
  year={2020},
  volume={},
  number={},
  pages={1-6}}

@misc{khatib2026llmedgerepo,
author = {Khatib, Maysam and others},
title = {{LLM Edge Continuum Benchmarking Dataset and Analysis}},
year = {2026},
howpublished = {\url{https://github.com/UCY-LINC-LAB/Edge-LLM-Inference-Benchmark}
}
}

@misc{llamacpp_quantize_readme,
  author       = {{ggml-org contributors}},
  title        = {{llama.cpp quantize README}},
  year         = {2026},
  howpublished = {\url{https://github.com/ggml-org/llama.cpp/blob/master/tools/quantize/README.md}},
}

@misc{prometheus,
  author       = {{Prometheus Authors}},
  title        = {Prometheus: Open Source Metrics and Monitoring for Your Systems and Services},
  year         = {2026},
  howpublished = {\url{https://prometheus.io/}},
  note         = {Accessed: 2026-03-13}
}

@misc{meross,
  author       = {{Meross Technology Limited}},
  title        = {Smart Plug for Google Home - Meross MSS310EU},
  howpublished = {\url{https://www.meross.com/en-gc/smart-plug/smart-plug-google-home/6}},
  note         = {Accessed: 2026-03-13},
  year         = {2026}
}

@misc{huggingface,
  author       = {{Hugging Face}},
  title        = {Hugging Face: The AI Community Building the Future},
  year         = {2026},
  howpublished = {\url{https://huggingface.co/}},
  note         = {Accessed: 2026-03-13}
}

@misc{ollama,
  author       = {{Ollama Inc.}},
  title        = {Ollama},
  year         = {2026},
  howpublished = {\url{https://ollama.com/}},
  note         = {Accessed: 2026-03-13}
}

@INPROCEEDINGS{mlcommons2025power,
  author={Arya Tschand and others},
  booktitle={2025 IEEE International Symposium on High Performance Computer Architecture (HPCA)}, 
  title={{MLPerf Power: Benchmarking the Energy Efficiency of Machine Learning Systems from $\mu$Watts to MWatts for Sustainable AI}}, 
  year={2025},
  volume={},
  number={},
  pages={1201-1216}}

@article{jiang2026edge,
  title={Edge large language models: a comprehensive survey},
  author={Jiang, Shan and Zhou, Xuecheng and Zhang, Mingjin and Xu, Changfu and Liao, Guocheng and Chen, Jianguo and Cao, Jiannong},
  journal={CCF Trans. Pervasive Comp. Interact.},
  pages={181--210},
  year={2026},
  publisher={Springer}
}

@INPROCEEDINGS{Mohammed2020,
  author={Mohammed, Thaha and Joe-Wong, Carlee and Babbar, Rohit and Francesco, Mario Di},
  booktitle={IEEE INFOCOM 2020 - IEEE Conference on Computer Communications}, 
  title={Distributed Inference Acceleration with Adaptive DNN Partitioning and Offloading}, 
  year={2020},
  volume={},
  number={},
  pages={854-863},
  publisher = {IEEE},
  address = {Piscataway, NJ}}

@INPROCEEDINGS{Kasioulis2024,
  author={Kasioulis, Michalis and Symeonides, Moysis and Ioannou, Giorgos and Pallis, George and Dikaiakos, Marios D.},
  booktitle={2024 IEEE International Conference on Cloud Engineering (IC2E)}, 
  title={Energy modeling of inference workloads with AI accelerators at the Edge: A benchmarking study}, 
  year={2024},
  volume={},
  number={},
  pages={189-196},
  publisher = {IEEE},
  address = {Piscataway, NJ}}

@ARTICLE{Morabito2025,
  author={Morabito, Roberto and Jang, SiYoung},
  journal={IEEE Internet Computing}, 
  title={Smaller, Smarter, Closer: The Edge of Collaborative Generative Artificial Intelligence}, 
  year={2025},
  volume={29},
  number={4},
  pages={7-15}}

@inproceedings {yu2022orca,
author = {Gyeong-In Yu and Joo Seong Jeong and Geon-Woo Kim and Soojeong Kim and Byung-Gon Chun},
title = {Orca: A Distributed Serving System for {Transformer-Based} Generative Models},
booktitle = {16th USENIX OSDI},
year = {2022},
isbn = {978-1-939133-28-1},
address = {Carlsbad, CA},
pages = {521--538},
publisher = {USENIX Association},
month = jul
}

@inproceedings{kwon2023pagedattention,
author = {Kwon, Woosuk and Li, Zhuohan and Zhuang, Siyuan and Sheng, Ying and Zheng, Lianmin and Yu, Cody Hao and Gonzalez, Joseph and Zhang, Hao and Stoica, Ion},
title = {Efficient Memory Management for Large Language Model Serving with PagedAttention},
year = {2023},
isbn = {9798400702297},
publisher = {Association for Computing Machinery},
address = {New York, NY, USA},
booktitle = {Proceedings of the 29th Symposium on Operating Systems Principles},
pages = {611–626},
numpages = {16},
location = {Koblenz, Germany},
series = {SOSP '23}
}

@inproceedings{agrawal2024sarathi,
author = {Agrawal, Amey and Kedia, Nitin and Panwar, Ashish and Mohan, Jayashree and Kwatra, Nipun and Gulavani, Bhargav S. and Tumanov, Alexey and Ramjee, Ramachandran},
title = {Taming throughput-latency tradeoff in LLM inference with sarathi-serve},
year = {2024},
publisher = {USENIX Association},
address = {USA},
booktitle = {Proceedings of the 18th OSDI},
articleno = {7},
numpages = {18},
location = {Santa Clara, CA, USA}
}

@inproceedings{georgiou2025,
  author={Georgiou, Joanna and Symeonides, Moysis and Pallis, George and Dikaiakos, Marios D.},
  booktitle={2025 IEEE International Conference on Edge Computing and Communications (EDGE)}, 
  title={Automating Multi-Tenancy Performance Evaluation on Edge Compute Nodes}, 
  year={2025},
  volume={},
  number={},
  pages={103-114},
  publisher = {IEEE},
  address = {Piscataway, NJ}
}

@inproceedings{xiao2023smoothquant,
author = {Xiao, Guangxuan and Lin, Ji and Seznec, Mickael and Wu, Hao and Demouth, Julien and Han, Song},
title = {SmoothQuant: accurate and efficient post-training quantization for large language models},
year = {2023},
booktitle = {Proceedings of the 40th ICML},
series = {ICML'23}
}

@article{lin2024awq,
author = {Lin, Ji and Tang, Jiaming and Tang, Haotian and Yang, Shang and Xiao, Guangxuan and Han, Song},
title = {{AWQ: Activation-aware Weight Quantization for On-Device LLM Compression and Acceleration}},
year = {2025},
issue_date = {December 2024},
publisher = {Association for Computing Machinery},
address = {New York, NY, USA},
volume = {28},
number = {4},
issn = {2375-0529},
journal = {GetMobile: Mobile Comp. and Comm.},
month = jan,
pages = {12–17},
numpages = {6}
}

@inproceedings{tian2025clone,
author = {Tian, Chunlin and Qin, Xinpeng and Tam, Kahou and Li, Li and Wang, Zijian and Zhao, Yuanzhe and Zhang, Minglei and Xu, Chengzhong},
title = {CLONE: customizing LLMs for efficient latency-aware inference at the edge},
year = {2025},
isbn = {978-1-939133-48-9},
publisher = {USENIX Association},
address = {USA},
booktitle = {Proceedings of the 2025 USENIX Conference on Usenix Annual Technical Conference},
articleno = {34},
numpages = {23},
location = {Boston, MA, USA},
series = {USENIX ATC '25}
}

@article{
hendrycks2021measuring,
  title={Measuring Massive Multitask Language Understanding},
  author={Dan Hendrycks and Collin Burns and Steven Basart and Andy Zou and Mantas Mazeika and Dawn Song and Jacob Steinhardt},
  journal={Proceedings of the International Conference on Learning Representations (ICLR)},
  year={2021}
}
\end{document}